\pdfoutput=1
\documentclass[review]{elsarticle}
\usepackage{lineno,hyperref}
\usepackage{booktabs}
\usepackage{siunitx}
\usepackage{graphicx}
\usepackage{placeins}
\usepackage{multirow}

\usepackage{natbib}
\setcitestyle{authoryear,open={(},close={)}}

\modulolinenumbers[5]
\journal{Advances in Space Research}

\newcommand{\me}{Mini-EUSO}
\newcommand{\mefovdeg}{\SI{44}{\degree}}

\newcommand{\meresolution}{\SI{6.11}{\kilo\metre}}
\newcommand{\megtu}{\SI{2.5}{\micro\second}}
\newcommand{\melensdiameter}{\SI{25}{\cm}}

\begin{document}
\begin{frontmatter}
\title{{\me}: A high resolution detector for the study of terrestrial and cosmic UV emission from the International Space Station}

\cortext[mycorrespondingauthor]{Corresponding author}
\author[sweden1,sweden2]{Francesca Capel\corref{mycorrespondingauthor}}
\ead{capel@kth.se}

\author[russia1,russia2]{Alexander Belov}
\author[japan,italy]{Marco Casolino}
\author[russia1]{Pavel~Klimov}

\author{for the JEM-EUSO Collaboration}

\address[sweden1]{Department of Physics, KTH Royal Institute of Technology, SE-106 91 Stockholm, Sweden}
\address[sweden2]{The Oskar Klein Centre for Cosmoparticle Physics, SE-106 91 Stockholm, Sweden}
\address[russia1]{D.V. Skobeltsyn Institute of Nuclear Physics, M.V. Lomonosov Moscow State University, 1(2), Leninskie Gory, 119991, Russia}
\address[russia2]{Faculty of Physics, M.V. Lomonosov Moscow State University, 1(2), Leninskie Gory, 119991, Russia}
\address[japan]{RIKEN, Hirosawa 2-1, Wako-shi, Saitama 351-01, Japan}
\address[italy]{Istituto Nazionale di Fisica Nucleare - Sezione di Roma Tor Vergata, Via Carnevale Emanuele, 00173, Italy}

\vspace{-0.5cm}

\begin{abstract}
The {\me} instrument is a UV telescope to be placed inside the International Space Station (ISS), looking down on the Earth from a nadir-facing window in the Russian Zvezda module. {\me} will map the earth in the UV range (300~-~400~\si{\nano\metre}) with a spatial resolution of {\meresolution} and a temporal resolution of \SI{2.5}{\micro\second}, offering the opportunity to study a variety of atmospheric events such as transient luminous events (TLEs) and meteors, as well as searching for strange quark matter and bioluminescence. Furthermore, {\me} will be used to detect space debris to verify the possibility of using a EUSO-class telescope in combination with a high energy laser for space debris remediation. The high-resolution mapping of the UV emissions from Earth orbit allows Mini-EUSO to serve as a pathfinder for the study of Extreme Energy Cosmic Rays (EECRs) from space by the JEM-EUSO collaboration. 
\end{abstract}


\end{frontmatter}

\section{Introduction: The EUSO program}
At the far end of the cosmic ray energy spectrum, with energies above \SI{50}{\exa\electronvolt}, lie the elusive Extreme Energy Cosmic Rays (EECRs). At such energies, the flux is as low as 1 particle/\si{\square\kilo\metre}/century and the effective area that can be observed by a detector is a key feature. The JEM-EUSO Collaboration aims to detect the UV light produced by EECR-induced extensive air showers (EAS) from the vantage point of low Earth orbit, thereby largely increasing the effective detector volume \citep{Collaboration:2015fz, Olinto15}. In order to lay the groundwork for such unprecedented observations, the JEM-EUSO collaboration has successfully initiated several pathfinder experiments. {\me}, along with EUSO-TA \citep{Kawasaki15}, EUSO-Balloon \citep{Scotti16}, and EUSO-SPB \citep{Wiencke16}, forms the next step towards the observation of EECRs from space. Additionally, UV nightglow measurements with a similar resolution to that of {\me} are currently being conducted by the TUS experiment on board the Lomonosov satellite \citep{TUS}. 

Previous relevant studies include NIGHTGLOW \citep{Barbier:2005ky} and satellite-borne Tatiana \citep{Garipov:2005ga} experiments. NIGHTGLOW was a balloon-borne UV telescope flying at an altitude of $\sim$~30~\si{\kilo\metre} for a duration of 8 hours, 4 days after the new moon. The mounted telescope was able to rotate, allowing measurements over a range of angles from nadir to 45\si{\degree} to the zenith. The Tatiana satellite carried a UV detector and operated in a polar orbit at 950~\si{\kilo\metre} for a duration of $\sim$~2~years. The detector was a single photomultiplier tube with a wide field of view ($\sim$~\SI{15}{\degree}) and without spatial resolution. The results from both experiments are consistent, reporting that the UV emission from Earth falls in the range of $3 \times 10^{11}$~-~$10^{12}$~photons~$\cdot$~\si{\per\square\metre\per\second\per\steradian}. The UV radiation measured by satellite missions is highly variable due to the presence of clouds, cities, aurora and other factors in the moving field of view. Both {\me} and TUS will build on these results with higher resolution measurements, as discussed in section \ref{sec:science}.

\begin{figure}[h]
	\begin{center}
		{\includegraphics[width=0.7\textwidth]{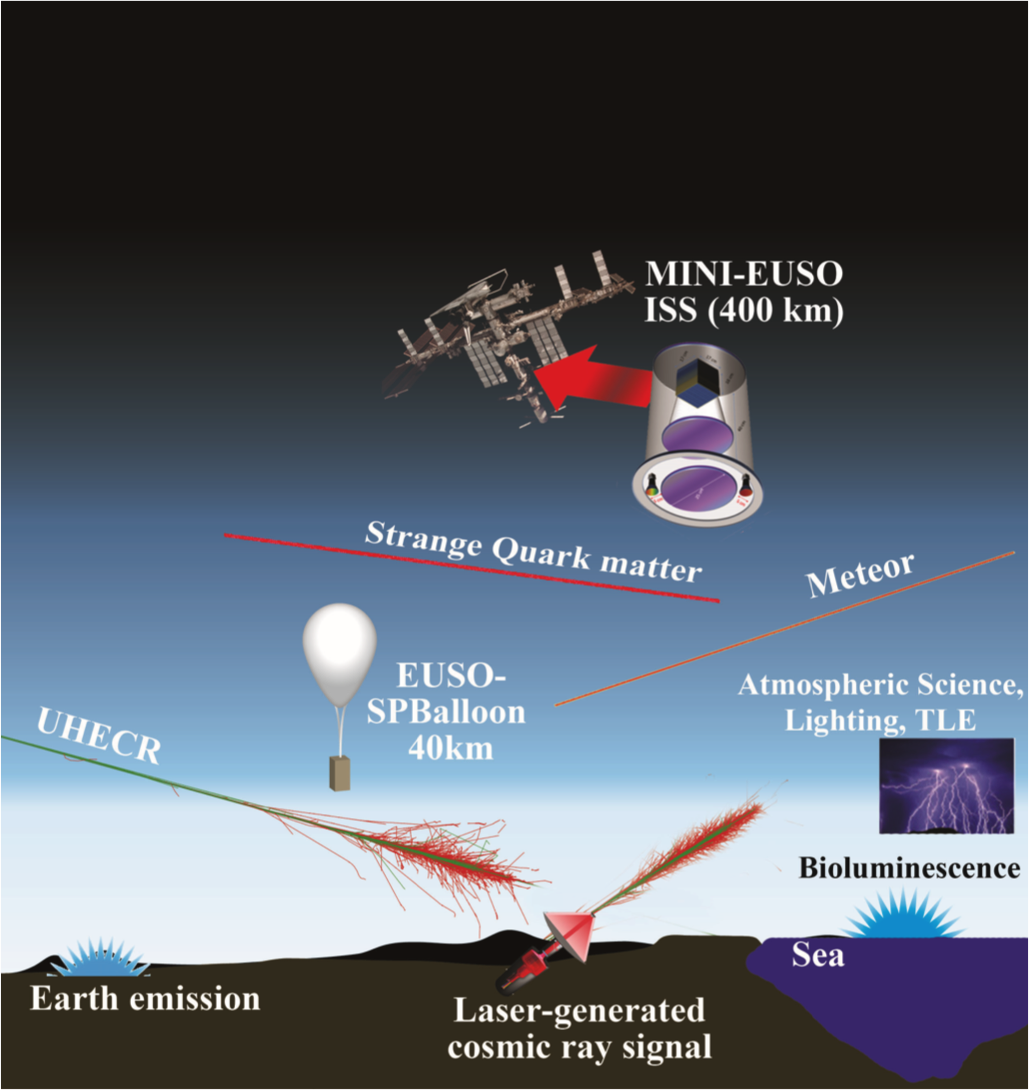}}
		\caption{ The {\me} mission summarised in one diagram. From the ISS, {\me} will observe a variety of interesting phenomena in the UV range, in addition to creating a high resolution UV map of the Earth.}
		\label{fig:MEmission}
	\end{center}
\end{figure}

The main goal of {\me} is to measure the UV emissions from Earth orbit. These observations will provide interesting data for the scientific study of a variety of UV phenomena such as transient luminous events (TLEs), meteors, space debris, strange quark matter (SQM) and bioluminescence, as summarised in Figure \ref{fig:MEmission}. Moreover, this will allow the characterisation of the UV emission level, which is essential for the optimisation of the design of future EUSO instruments for EECR detection. {\me} is approved as a joint project by the Italian (ASI) and Russian (Roscosmos) space agencies and is included in the long-term program of space experiments on the Russian segment of the ISS under the name ``UV-Atmosphere''. It is scheduled to be launched to the ISS, where it will be placed at a nadir-facing, UV-transparent window on the Russian Zvezda module. The integration of the instrument is currently at an advanced stage in order to be compliant with a launch opportunity in late 2017 to early 2018.

\section{Instrument overview}
\label{sec:overview}
{\me} is based on one EUSO detection unit, referred to as the Photo Detector Module (PDM). The PDM consists of 36 multi-anode photomultiplier tubes (MAPMTs), each with 64 pixels, for a total of 2304 pixels. The MAPMTs are provided by Hamamatsu Photonics, model R11265-M64, and covered with a 2~\si{\milli\metre} BG3 UV filter with anti-reflective coating. The full {\me} telescope is made up of 3 main systems, the optical system, the PDM and the data acquisition system. The optical system of 2 Fresnel lenses is used to focus light onto the PDM in order to achieve a large field of view ({\mefovdeg}) with a relatively light and compact design, well-suited for space application. The PDM detects UV photons and is read out by the data acquisition system with a sampling rate of \SI{2.5}{\micro\second} and a spatial resolution of {\meresolution}. The key parameters of the instrument are summarised in Table \ref{table:MEparams}.

\begin{figure}[h]
	\begin{center}
		{\includegraphics[width=\textwidth]{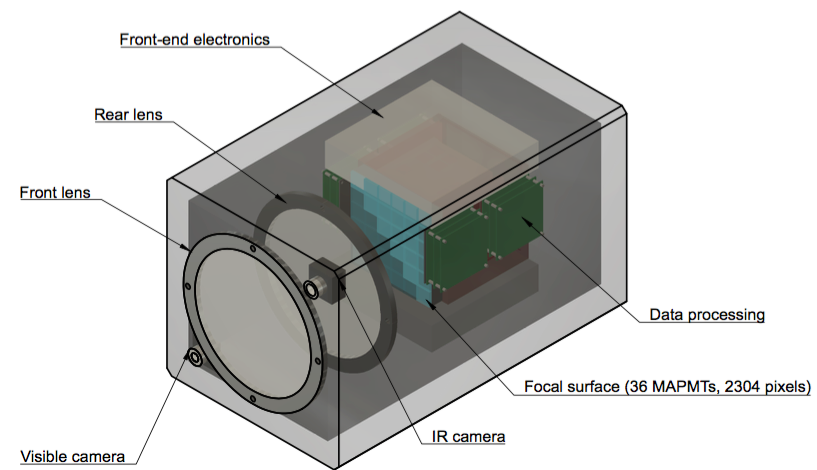}}
		\caption{ {\me} conceptual design. The optical system with two double sided Fresnel lenses ({\melensdiameter} diameter) focuses the UV light on to a focal surface consisting of a single PDM, made up of 36 MAPMTs and 2304 pixels. Ancillary detectors are the visible and near infra-red cameras. The instrument dimensions are 37 $\times$ 37 $\times$ 62 \si{\cubic\centi\metre}.}
		\label{fig:me_highlevel}
	\end{center}
\end{figure}

In addition to the main detector, {\me} contains two ancillary cameras for complementary measurements in the near infrared (NIR from 1500 to 1600~\si{\nano\metre}) and visible (VIS from 400 to 780~\si{\nano\metre}) range. Both cameras are produced by PointGrey. The NIR is a Chameleon 1.3 MP Mono USB 2.0 (CCD: Sony ICX445) coated with phosphorous as an infrared filter, with a resolution of $1296 \times 964$ and a maximum frame rate of 18 fps \citep{ptgrey:NIR}. The VIS is a Firefly MV 1.3 MP Color USB 2.0 (CMOS: Sony IMX035) with a resolution of $1328 \times 1048$ and a maximum frame rate of 23 fps \citep{ptgrey:VIS}. A fixed \SI{8.5}{\milli\metre} focal lens provided by Edmund Optics is fitted to each camera. These cameras are placed outside the optical system and acquire data independently of the PDM. The main task of the cameras is to provide atmospheric monitoring in order to better understand the UV luminosity measurements. The complete {\me} instrument is contained in a box with a connection to ISS for power/grounding and the interface to the UV-transparent window, from which Mini-EUSO will observe in a fixed position facing the nadir direction. Figure \ref{fig:me_highlevel} shows the layout of the main systems. 

\begin{table}[h]
\centering
\caption{Parameters of the {\me} instrument, including the field of view (FoV) for each independent pixel. Parameters for the proposed JEM-EUSO instrument are also shown \citep{Collaboration:2015fz}, along with the EUSO-SPB pathfinder and TUS for comparison \citep{Wiencke16, TUS}. {\me} is defined as a pathfinder for the JEM-EUSO mission and as such, the FoV and pixel size for {\me} were chosen to give a comparable UV signal and exposure. The spatial resolution given is the extent of a single pixel on the Earth's surface for a nadir-pointing instrument. Pixel size on the MAPMT for all EUSO instruments is a circle of \SI{3.3}{\milli\metre} diameter, whereas for TUS the pixel is a single circular PMT of diameter \SI{13}{\milli\metre}.}
\vspace{3mm}
  \begin{tabular}{ ccccc }
    \toprule
  & \textbf{{{\me}}} & \textbf{JEM-EUSO} & \textbf{EUSO-SPB} & \textbf{TUS} \\ \midrule
     \textbf{Spatial} &  \multirow{2}{*}[0em]{ \SI{5}{\kilo\metre} }& \multirow{2}{*}[0em]{\SI{560}{\metre}} & 
     \multirow{2}{*}[0em]{\SI{130}{\metre}}&\multirow{2}{*}[0em]{\SI{5}{\kilo\metre}} \\
     \textbf{resolution} & & & & \\
     \textbf{Temporal} & \multirow{2}{*}[0em]{\SI{2.5}{\micro\second}} & \multirow{2}{*}[0em]{\SI{2.5}{\micro\second}} &
     \multirow{2}{*}[0em]{\SI{2.5}{\micro\second}} & \multirow{2}{*}[0em]{\SI{800}{\nano\second}} \\ 
     \textbf{resolution} & & & & \\ \midrule
     \textbf{Aperture} & \multirow{2}{*}[0em]{Circular} & \multirow{2}{*}[0em]{Circular} & 
     \multirow{2}{*}[0em]{Square} & Hexagonal \\
     \textbf{shape} & & & & segments \\
     \textbf{Aperture} & \multirow{2}{*}[0em]{490~\si{\square\cm}} & \multirow{2}{*}[0em]{\SI{4.5e4}{\cm\squared}} &
     \multirow{2}{*}[0em]{\SI{1e4}{\cm\squared}} & \multirow{2}{*}[0em]{\SI{2e4}{\cm\squared}} \\
     \textbf{area} & & & & \\ \midrule
      \textbf{FoV} & {\mefovdeg} & 60\si{\degree} & 11\si{\degree} &  9\si{\degree}\\
     \textbf{FoV/Pixel} & 0.8\si{\degree} & 0.08\si{\degree} & 0.23\si{\degree} &  0.8\si{\degree}\\
   \textbf{N$\mathbf{^o}$ pixels} & \SI{2304} & \SI{315648} & 2304 & 256 \\
        \bottomrule
  \end{tabular}
  \label{table:MEparams}
\end{table}

\begin{figure}[ht]
	\begin{center}
		{\includegraphics[width=\textwidth]{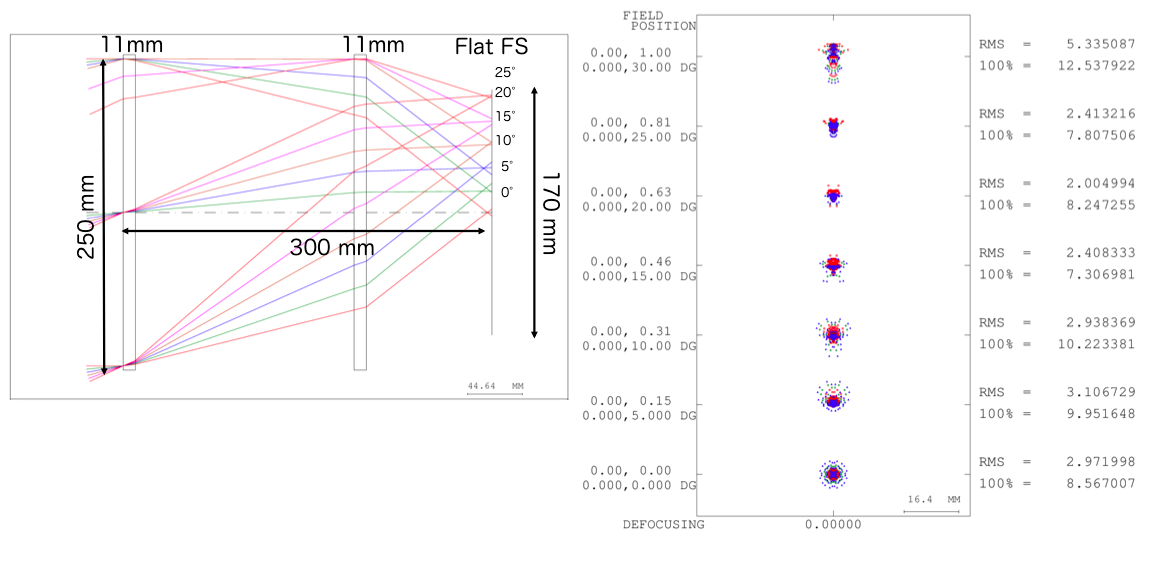}}
		\caption{The left figure shows the lens design of {\me} configuration. Key dimensions are shown in bold text. The lenses are \SI{11}{\milli\metre} thick, the focal length is \SI{300}{\milli\metre} and the FS (focal surface) is a square of side \SI{170}{\milli\metre}. In the figure on the right, the PSF is shown for a range of field positions. Incoming light is parallel with different inclinations as shown on the left of the figure. The RMS and 100\% values are given in units of~\si{\mm}. }
		\label{fig:optDesign}
	\end{center}
\end{figure}

\begin{figure}[ht]
	\begin{center}
		\includegraphics[width=0.7\textwidth]{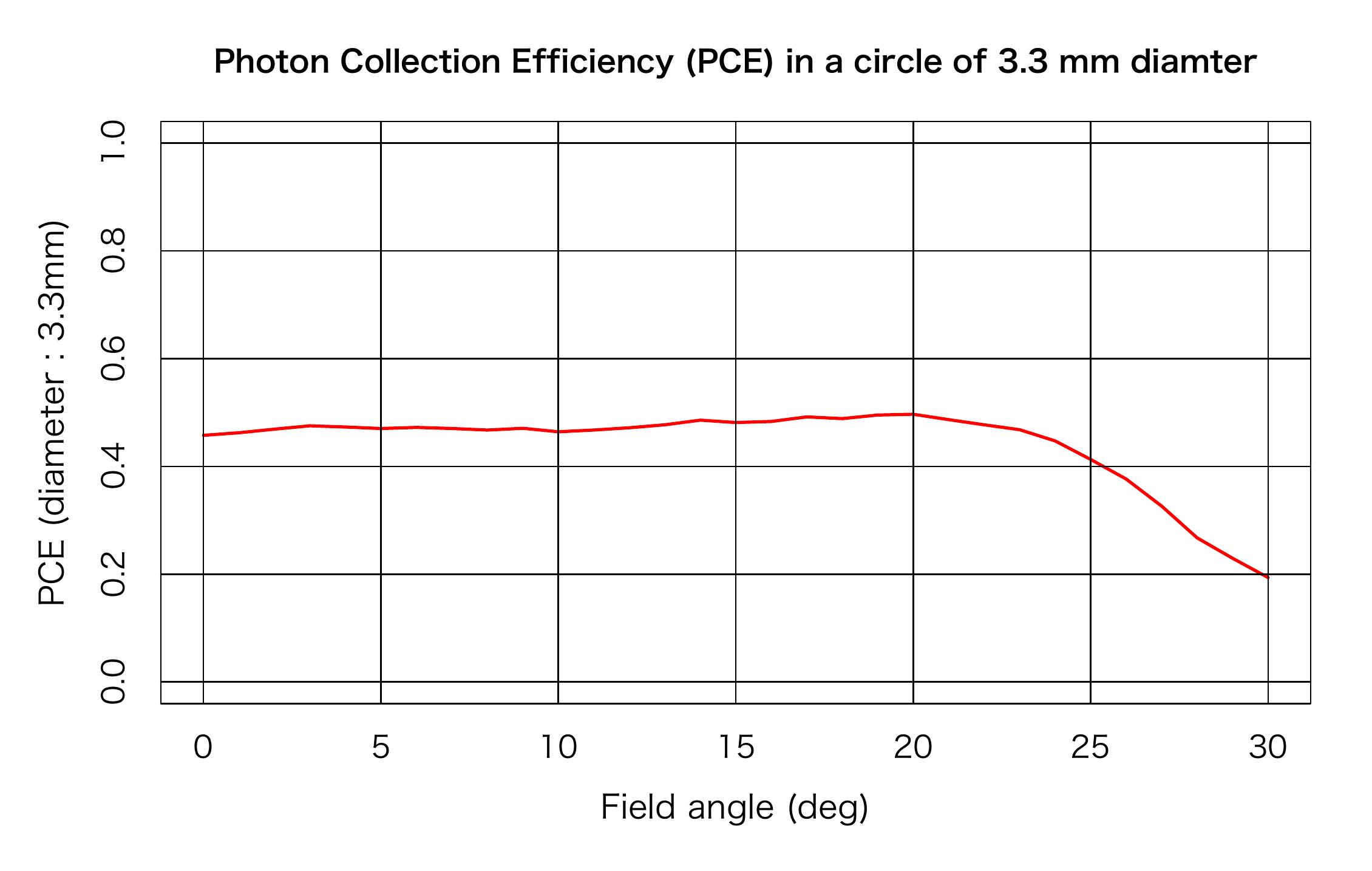}
		\includegraphics[width=0.7\textwidth]{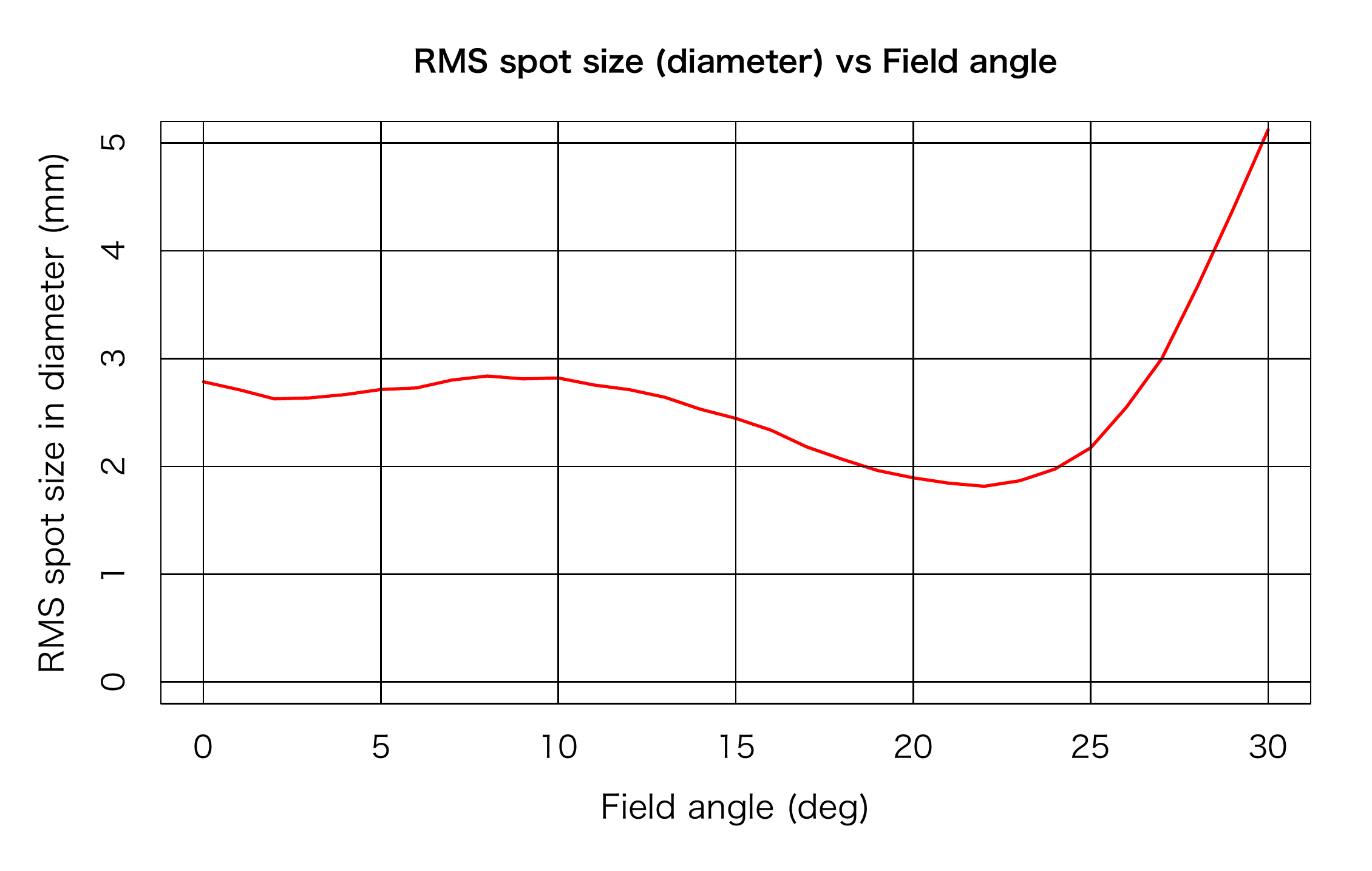}
		\caption{Top: The photon collection efficiency in 1 pixel (a circle of diameter \SI{3.3}{\milli\metre}) as a function of the angle at which photons enter the first lens. Bottom: The RMS spot size in \si{\milli\metre}, also as a function of field angle.}
		\label{fig:optDesign_sim}
	\end{center}
\end{figure}

The optical system of {\me} is composed of two double-sided Fresnel lenses. Each lens is made of PMMA, {\melensdiameter} in diameter, \SI{11}{\mm} thick and weighing \SI{0.8}{\kg}. The effective focal length is \SI{300}{\mm} and the field of view is {\mefovdeg}. Ray tracing simulations have been developed to calculate the photon collection efficiency (PCE) of the optical system, defined as the number of photons which arrive in one pixel size divided by the number of photons incident upon the front lens. For the purpose of the ray tracing simulations, a single 2.9~$\times$~2.9~\si{\mm\squared} square pixel of the PMT is approximated as a circle of diameter \SI{3.3}{\milli\metre}, giving the equivalent circular area. Figure \ref{fig:optDesign} shows the optical system layout and the point spread function (PSF). The PCE is shown in Figure \ref{fig:optDesign_sim}, alongside the RMS spot size as a function of field angle. The results are achieved using a simulation run with 3 wavelengths of \SI{337}{\nano\metre}, \SI{357}{\nano\metre} and \SI{391}{\nano\metre}, each with equal intensity. This gives an approximation of the chromatic response over the accepted wavelength band of 300~-~400~\si{\nano\metre}, for the primary emission lines of EAS-induced air fluorescence \citep{Kakimoto:1996io}. Factors due to surface reflection, material absorption, surface roughness and Fresnel facet back cut are also taken into account. In addition to the optical system of Mini-EUSO, it is important to recall that Mini-EUSO will be positioned looking through a UV-transparent, nadir-facing window in the Russian Zvezda module. The transmission function of this window is fairly constant at a value of 86\% over a wide wavelength range, including the 300~-~400~\si{\nano\metre} band detected by Mini-EUSO. This is not included in the simulated instrument response as it will not affect the results, just slightly increase the thresholds.   

The array of MAPMTs in the {\me} PDM is powered by a Cockroft-Walton high voltage power supply (HVPS), allowing for a low power solution to providing the high voltage needed by the PMTs. In order to protect the MAPMTs from potentially damaging high current levels, the HVPS has a fast ($<$~\SI{3}{\micro\second}) built-in switch which drastically reduces the gain of the MAPMTs when the anode current surpasses a threshold value. The photon flux of the phenomena which will be observed by {\me} varies on the order of $10^6$, from background levels of 1~count/pixel/GTU (1 GTU = {\megtu}), to bright TLEs and meteors. The background value stated is the number of photons detected by {\me} (photon counts). This value obtained for JEM-EUSO by \citet{AdamsJr:2013bd}, scales roughly in the same way to {\me}, taking into account the ratio of the optics size and pixel field of view. The HVPS handles this large range via the implementation of a second switch, controlled by the Zynq board. This second switch works by reducing the gain in a controlled way, as the incident photon flux increases in order to cover the full dynamic range of $10^6$, without damage to the MAPMTs. The anode current is continuously monitored and higher gain is restored once the current has dropped beneath the threshold level.

\begin{figure}[ht]                                                                                                                                                                                                                                                                                                                                                                                                               
	\begin{center}
		\includegraphics[width=\textwidth]{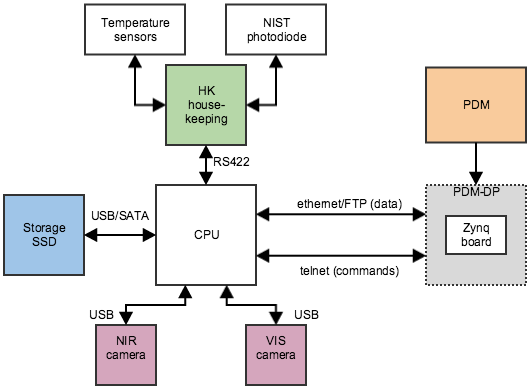}
		\caption{The data acquisition system of {\me} with the main interfaces shown. Incoming data from the PDM ASICs is triggered in the Zynq board and then passed to the CPU via an ethernet link. The CPU combines this data with that of the housekeeping system and the ancillary cameras. Data is then moved to onboard SSDs for storage.}
		\label{fig:dpdata}
	\end{center}
\end{figure}

The data acquisition system consists of the front-end electronics, the PDM-DP (PDM data processing) sub-system based on a Xilinx Zynq XC7Z030 \citep{zynq} system on chip and a PCIe/104 form factor CPU. This is an evolution of the system used in previous EUSO pathfinders, such as EUSO-TA, EUSO-Balloon and EUSO-SPB, incorporating the functionality of several subsystems into one board. Incoming photon pulses are pre-amplified and digitised by the SPACIROC3 ASICs \citep{BlinBondil:2014ve} at intervals of \SI{2.5}{\micro\second}, referred to as the Gate Timing Unit or GTU. The signal is then triggered and time-stamped in the Zynq FPGA, before being passed to the CPU for data management and storage. The Zynq chip contains a Xilinx Kintex7 FPGA, with an embedded dual core ARM9 CPU processing system and is responsible for of the majority of the data handling including data buffering, configuration of the SPACIROC3 ASICs, triggering, synchronisation and interfacing with the separate CPU system. In addition, the high-voltage applied to the PMTs is also controlled here, allowing fast real-time response to high signal. The data acquisition system is summarised in Figure \ref{fig:dpdata}. The CPU performs the control of the instrument sub-systems as well as the data management and storage, housekeeping, switching between operational modes and collecting data from the NIR and VIS cameras. Data is stored on board in SSDs which are periodically returned to Earth from the ISS, as it is not possible to telemeter such a large amount of data. In order to monitor the status of the Mini-EUSO instrument, smaller ``quick-look'' data samples will be telemetered at regular intervals by the astronauts on board the ISS. The instrument has no direct connection to the ISS network, so time-stamping will be achieved using the on-board CPU clock, which is regularly synchronised with the Zynq FPGA. In addition to this, Mini-EUSO will make observations of a ground-based laser system (as described in Section~\ref{sec:science}) which will allow offline synchronisation of the on-board CPU and use of the publicly available ISS ephemeris data in the subsequent analysis (NORAD Two-Line Element sets, \citet{norad}).

A multi-level trigger system is implemented in the Zynq programmable logic in order to maximise the scientific output of the instrument. The motivation for this is to capture events of interest on short timescales whilst continuously imaging and mapping the UV emissions. EECR-like events are triggered with a resolution of \SI{2.5}{\micro\second} (L1 trigger), TLEs with a resolution of \SI{320}{\micro\second} (L2 trigger) and there is an additional continuous readout with a resolution of \SI{40.96}{\milli\second} (L3 data). For the L1 trigger logic, each pixel is considered as independent due to its large field of view at ground of {\meresolution}, thus photons traveling at the speed of light take $\sim$~20~\si{\micro\second} to cross one pixel. The background level is calculated over 128 GTU (1~GTU~$=$~\SI{2.5}{\micro\second}) and used to set a threshold of $8 \sigma$ over the background level. The 128 GTU of integrated data used to calculate the background level is also passed as an input to the L2 trigger. Signal in a single pixel is integrated over 8 GTU and when the threshold is surpassed, the event is triggered and the whole 128 GTU packet is stored, centred on the event. The L2 trigger functions analogously but instead takes an integrated input of 1 GTU$_{L2} =$ \SI{320}{\micro\second}. The L2 also passes 128 integrated GTU$_{L2} =$ 1 GTU$_{L3} =$ \SI{40.96}{\milli\second} to the next level of the trigger algorithm. Key parameters of the trigger algorithm, such as the threshold and integration period, are configurable and can be changed in-flight. Every \SI{5.24}{\second} the three data types are read out to the CPU for permanent storage along with housekeeping data. A more detailed description of the trigger algorithm and the PDM-DP system is given in \citet{Trigger}, (submitted).

The NIR and VIS cameras will operate with a trigger passed from the Mini-EUSO PDM in order to provide multi-wavelength measurements of slower atmospheric events, such as meteors and nuclearites. When not triggered, the cameras will operate continuously to provide complementary measurements on the atmospheric status at the time of measurement, matching the third level of data from the Mini-EUSO trigger.

\section{Mission objectives}
\subsection{Scientific objectives}
\label{sec:science}
The objective of {\me} is to perform, for the first time, high-resolution mapping of the emission from the night-Earth in the UV band (300 - 400~\si{\nano\metre}), in order to study the UV luminosity. Previous missions, for example the Tatiana experiment \citep{Garipov:2005ga}, with a spatial resolution of $\sim$~100 km, have found a minimum flux level of the order of $3 \times 10^{11}$~photons~$\cdot$~\si{\per\square\metre\per\second\per\steradian}. However, this is an estimate for the dark areas of the Earth during moonless nights, with values a factor of 2-5 higher being possible over clouds or cities and 1-2 orders of magnitude higher over aurora regions. Other balloon-borne experiments such as EUSO-Balloon and BABY \citep{Catalano:2002vu}, have also made higher resolution (of the order of 10 km) measurements of the UV emission over ground, but only in localised areas and at altitudes below 40 km, meaning they are unable to detect airglow emission, aurora or other high-altitude effects. NIGHTGLOW has made measurements with a resolution of $\sim$~3~\si{\kilo\metre} from an altitude of 30~\si{\kilo\metre} at a range of zenith angles from nadir to 45\si{\degree} off-zenith, allowing observation of the airglow emission. The results, as presented in \citet{Barbier:2005ky}, show that airglow emission can contribute to an increase in the UV emission level of a factor of $\sim$~2.6.  {\me} will observe with a temporal resolution of \SI{2.5}{\micro\second} and a spatial resolution of {\meresolution} and as such will be able to characterise the UV luminosity of the entire Earth with unprecedented detail. 

\begin{figure}
  \centering
  \includegraphics[width=0.75\textwidth]{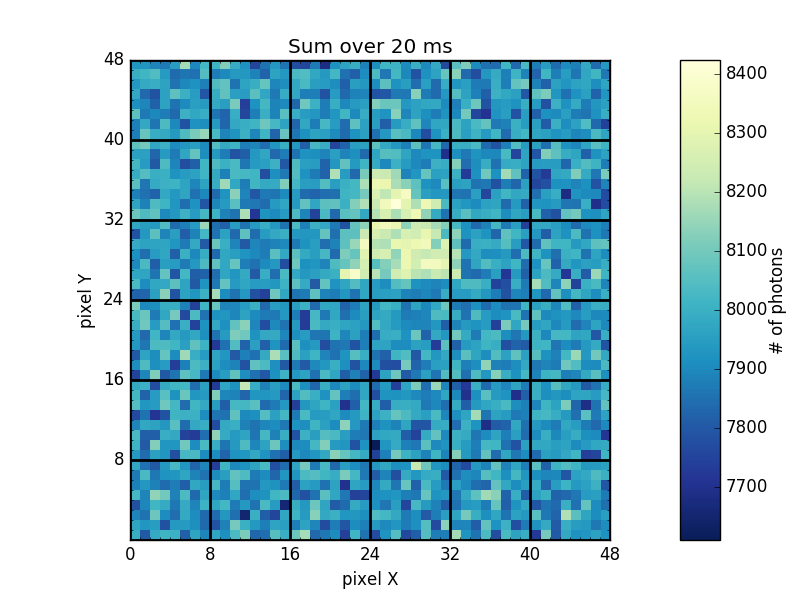}
  \includegraphics[width=0.75\textwidth]{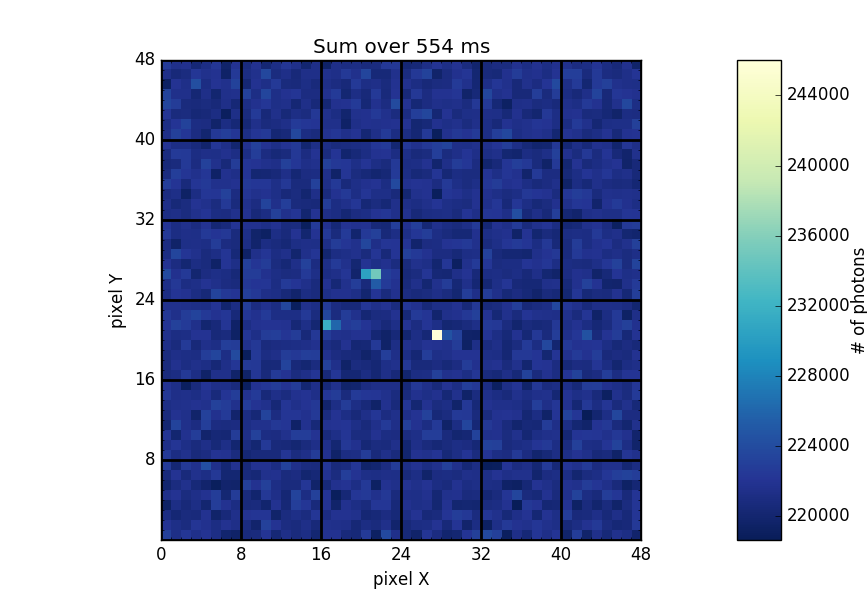} 
  \caption{Expected light track of a diffuse elf (top) and 3 localised blue jet events (bottom) as they would be detected by {\me}. Background emission is also included, centred on 1~photon~count/pixel/GTU \citep{AdamsJr:2013bd}.}
  \label{fig:tles}
 \end{figure}

TLEs such as blue jets, sprites and elves have been discovered relatively recently and are still not well understood. A detailed review of developments in the experimental and modeling studies of TLEs is presented in \citet{Pasko:2011ev}. These upper-atmospheric events are luminous in the UV and have high frequencies \citep{garipov2010program, panasyuk2010transient}, thus should be well characterised to avoid interference with EECR detection and triggering. {\me} has a dedicated trigger algorithm to capture TLEs and other millisecond scale phenomena at high resolution. This data could help improve the understanding of the formation mechanisms of these filamentary plasma structures, complementing atmospheric science experiments SMILES~\citep{randel1996isolation} and ASIM~\citep{neubert2009asim}. Figure~\ref{fig:tles} shows examples of typical TLEs as {\me} is expected to detect them (see Table \ref{table:tle_param} for the definition of typical TLE parameters). The HV switching system of {\me} will modify the detection efficiency of the MAPMTs by changing the voltage between cathode and the first dynodes, as described in Section~\ref{sec:overview}. In this way, the full dynamic range of {\me} spans over 6 orders of magnitude in photon flux and many different types of TLEs can be detected.

\begin{figure}
  \centering
  \includegraphics[width=0.75\textwidth]{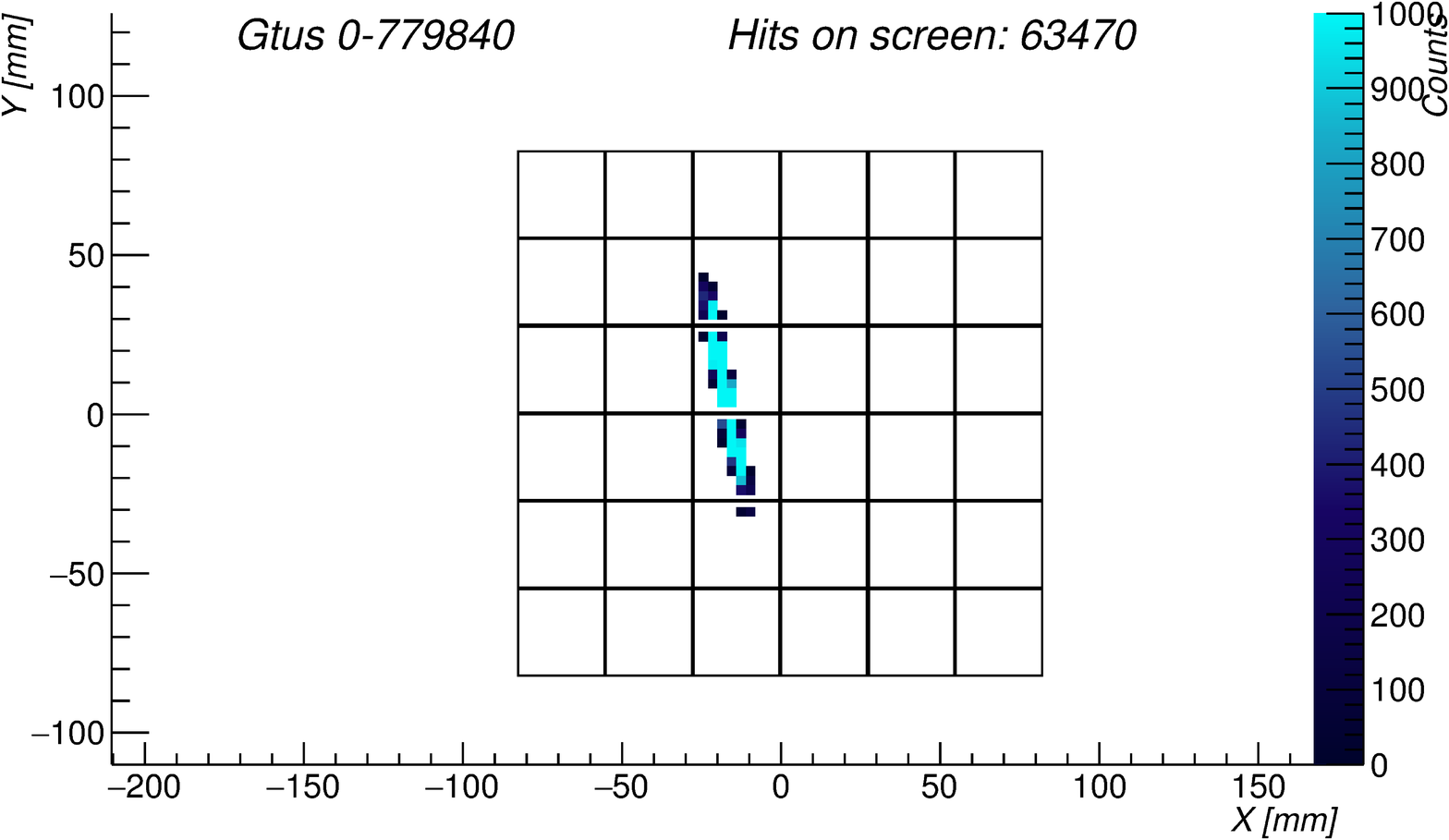}
  \includegraphics[width=0.75\textwidth]{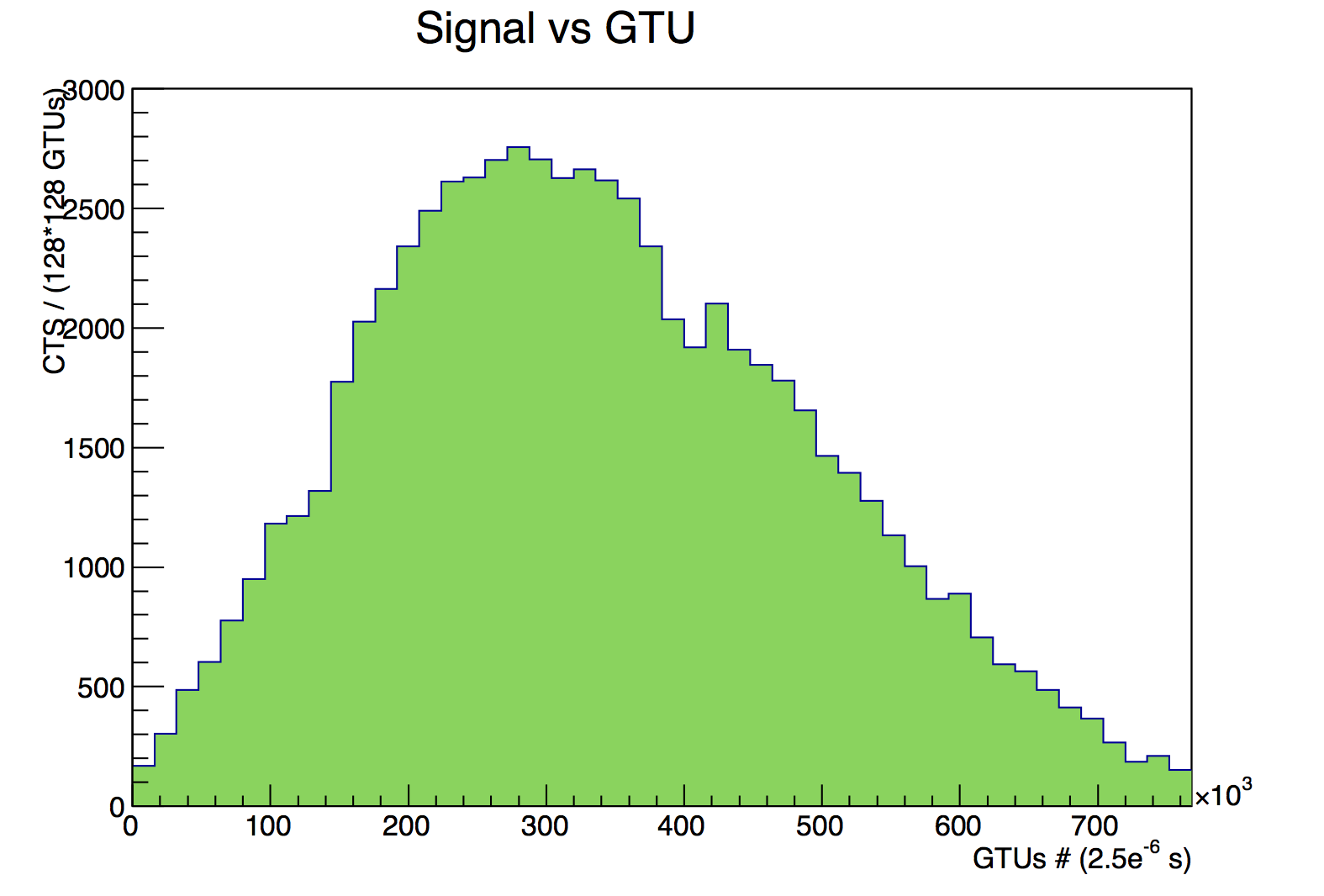}
  \caption{Top: Expected light track of a meteor of absolute magnitude $M=+5$ detected by Mini-EUSO (the effects of UV-nightglow are not included and a threshold has been applied at 30 counts). Bottom: Expected light profile. Each time bin on the x-axis corresponds to an integration time of \SI{40.96}{\milli\second}, the resolution of the level 3 data from {\me}. Figure taken from \citet{PSS}.}
  \label{fig:minieuso-meteor}
 \end{figure}

{\me} will also be able to see slower events such as meteors, fireballs, strange quark matter (SQM) and space debris with magnitudes of M~$<$~+5. In optimal dark conditions, the signal (integrated at steps of 40.96 ms) will exceed the UV-nightglow level by 3~-~4$\sigma$. These events will be detected using offline trigger algorithms on ground, although it is also possible that bright meteor events that appear suddenly can be triggered by the level 2 trigger, allowing higher resolution data for such events. Table~\ref{mag} shows the expected rate of meteors and intensity of the signal as a function of the magnitude. Figure~\ref{fig:minieuso-meteor} shows an example of a meteor track having absolute magnitude $M = +5$ crossing the field of view of {\me} with a 45$^\circ$ inclination with respect to the nadir axis. The meteor speed is \SI{70}{\kilo\metre\per\second} and its duration is \SI{2}{\second}.

\begin{table}
  \caption{Meteor emissions expected by {\me}. For a range of values of absolute magnitudes in visible light, the table lists the corresponding flux in the $U$-band of the Johnson-Morgan UBV photometric system~\citep{spitzer}, numbers of photons/\si{second} (assuming that the meteor is located at a height of \SI{100}{\kilo\metre} and is observed by the ISS in the nadir direction) and photo-electrons/\si{\milli\second} for {\me}. The typical mass of the meteor, and the number of events expected to be observed by {\me} (by assuming a duty cycle of $0.2$) are also shown. The relationship between mass and magnitude has been obtained following \citet{ra1968}. For more details see \citet{PSS}.}
\label{mag}
  \begin{center}
  \begin{tabular}{cccccc}
    \hline
     \textbf{Abs.} & \textbf{U-band flux} & \textbf{photons} & \textbf{photo-e$^-$} & \textbf{mass} & \textbf{event} \\
 \textbf{mag}    & \textbf{(erg/s/cm$^2$/A)} & \textbf{(s$^{-1}$)} & \textbf{ms$^{-1}$ }& \textbf{(g)} & \textbf{rate} \\
    \hline
    +5 & 4.2$\cdot$10$^{-11}$ &  2.7$\cdot$10$^ {8}$ & 10$^2$ & 10$^{-2}$ & 2.4/min   \\
    0 & 4.2$\cdot$10$^{-9}$  &  2.7$\cdot$10$^{10}$ & 10$^4$ & 1 & 0.11/orbit\\
   -5 & 4.2$\cdot$10$^{-7}$  &  2.7$\cdot$10$^{12}$ & 10$^6$ & 100 & 2.5/year  \\
    \hline
  \end{tabular}
  \end{center}
\end{table}

\begin{figure}
	\begin{center}
		\includegraphics[width=0.75\textwidth]{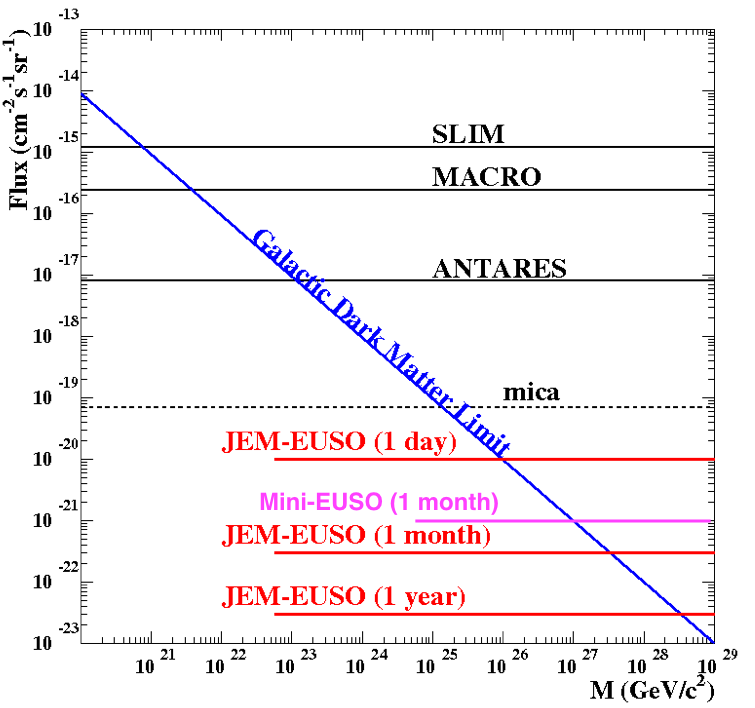}
		\caption{The 90\% confidence level upper limit to SQM with {\me} after 1 month of observation as compared to the limits set by other experiments: MACRO \citep{Ambrosio:449562}, SLIM \citep{Sahnoun:2009fr}, ANTARES \citep{Pavalas:2013cr} and MICA \citep{Price:1988ge}. The limits achieved by the planned JEM-EUSO instrument are also shown, these are stronger as JEM-EUSO has an aperture of around 4 times the size of {\me}, as well as a higher duty cycle and lower energy threshold.}
		\label{fig:sqm}
	\end{center}
\end{figure}

Even after just 1 month of observation at a minimal UV-nightglow level, {\me} will be able to set a new upper limit on the detection of SQM, as shown in Figure \ref{fig:sqm}. SQM is composed of roughly equal numbers of up, down and strange quarks, and can form stable macroscopic nuggets referred to as nuclearites \citep{DeRujula:1984axn}. As described in \citet{AdamsJr:2014cp}, these nuclearites create a UV signal upon interaction with the atmosphere which can be detected by Mini-EUSO. The nuclearite signal is easily discerned from meteor tracks as we expect much higher velocities on the order of $\sim$~100~\si{\kilo\metre\per\second}, compared to a maximum of around \SI{72}{\kilo\metre\per\second} for meteors. Figure \ref{fig:sqm} shows the upper limit on the nuclearite flux of $10^{-21}$~\si{\per\square\centi\metre\per\steradian\per\second} for a null detection of nuclearites based on the conservative assumption that events with a projected velocity below \SI{190}{\kilo\metre\per\second} are rejected.

\begin{figure}
  \centering
  \includegraphics[width=0.75\textwidth]{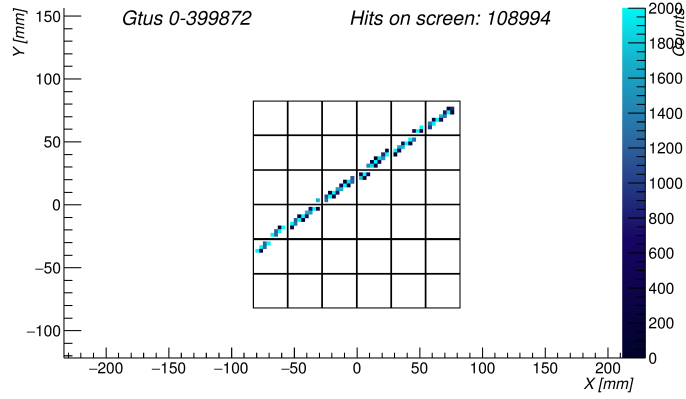}
  \caption{Expected light track of a piece of space debris flying at a relative speed of \SI{10}{\kilo\metre\per\second} at an altitude of \SI{10}{\kilo\metre} below the ISS whose signal is $\sim$~100 counts/ms on the {\me} focal surface. The track is followed for \SI{1}{\second} (the effects of UV-nightglow are not included and a threshold has been applied at 30 counts).}
  \label{fig:debris}
 \end{figure}

The observation of space debris is also a highly relevant issue. {\me} is effectively a high-speed camera with a large field of view and will be used as a prototype for the detection of space debris during the twilight periods of observation (when debris are illuminated by the sun, but the instrument is in darkness). In the future, larger scale EUSO experiments could be used in conjunction with a novel high efficiency fibre-based laser system (CAN) to provide a space-based debris remediation system \citep{Ebisuzaki:2015fm}. Figure~\ref{fig:debris} shows an example of integrated track of a space debris flying at a speed of \SI{10}{\kilo\metre\per\second} at an altitude of 10 km below the ISS. The signal is 100~counts/ms on the {\me} focal surface and the track is followed for \SI{1}{\second}, during which the signal is assumed to be constant and no UV-nightglow light has been added. This is preliminary work that shows the potential of {\me} for debris detection, a more realistic implementation of the expected signal in {\me} as a function of the size of the debris, as well as its reflectance and illumination by the sun, is currently under development.

\begin{figure}[ht]
	\begin{center}
  		\includegraphics[width=\textwidth]{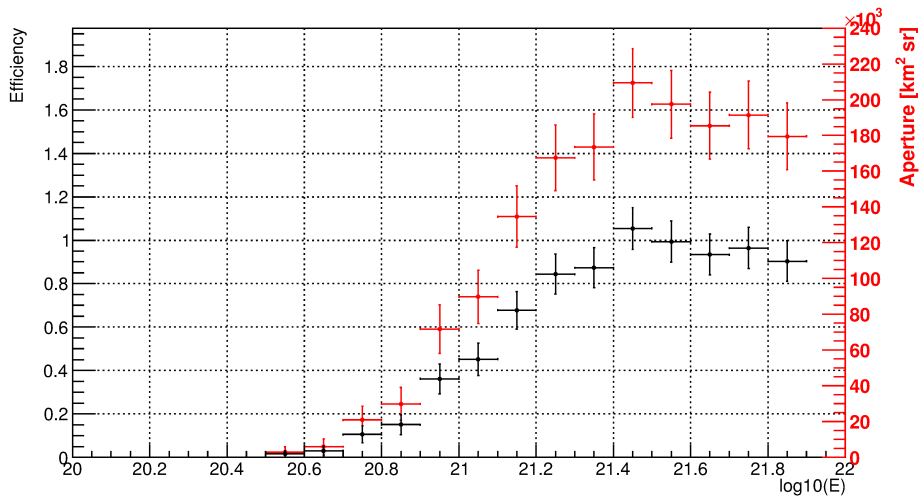}	
		\caption{The detection efficiency (on the left axis, in black) and geometric aperture, i.e. geometry factor, (on the right axis, in red) are shown as a function of the EAS energy, E, in \si{\electronvolt} . A UV background level of 1~photon~count/pixel/GTU \citep{AdamsJr:2013bd} was considered in both cases.}
		\label{fig:trigeff}
	\end{center}
\end{figure}

\begin{figure}[ht]
  \centering
  \includegraphics[width=0.75\textwidth]{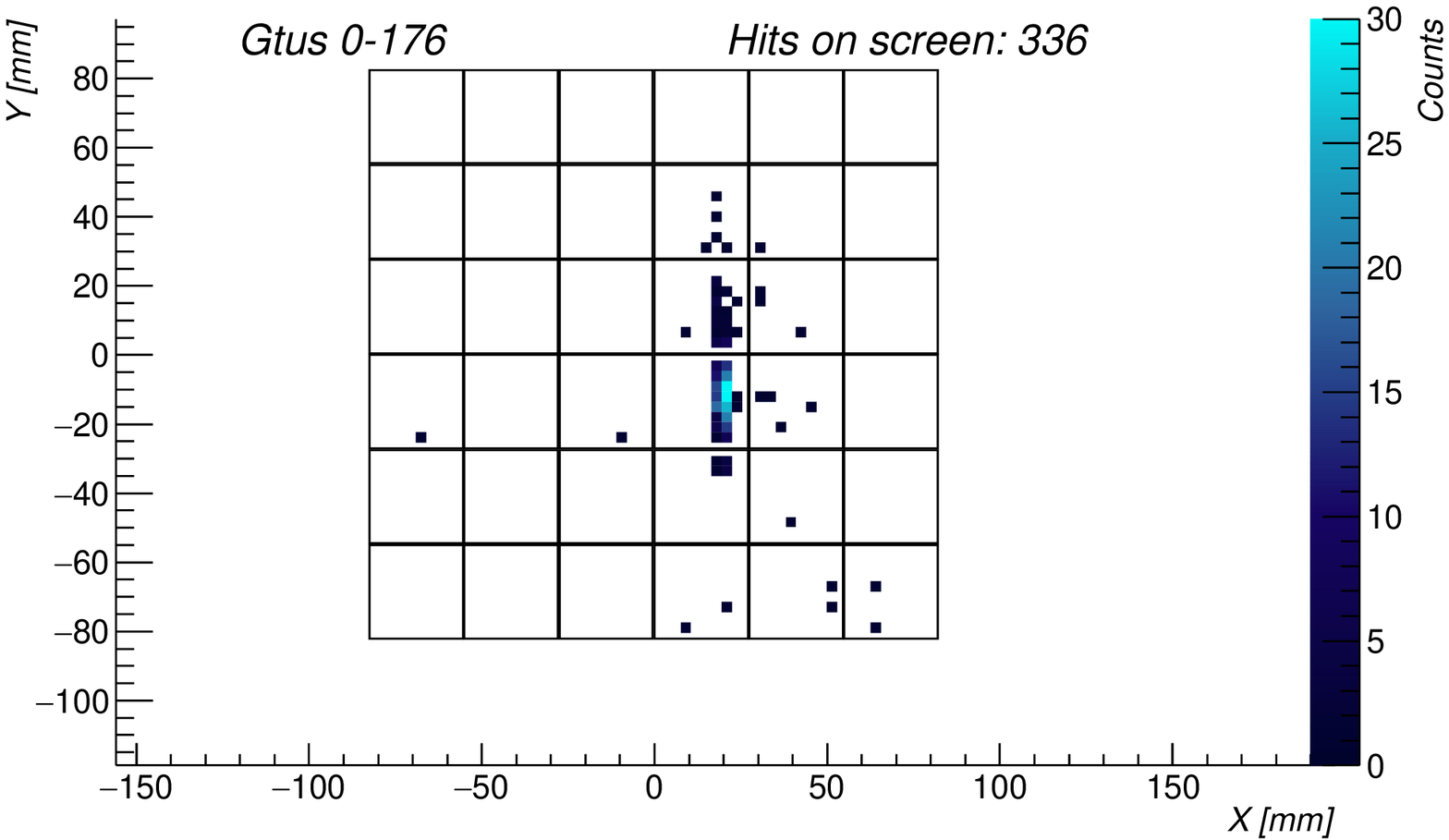}
  \includegraphics[width=0.75\textwidth]{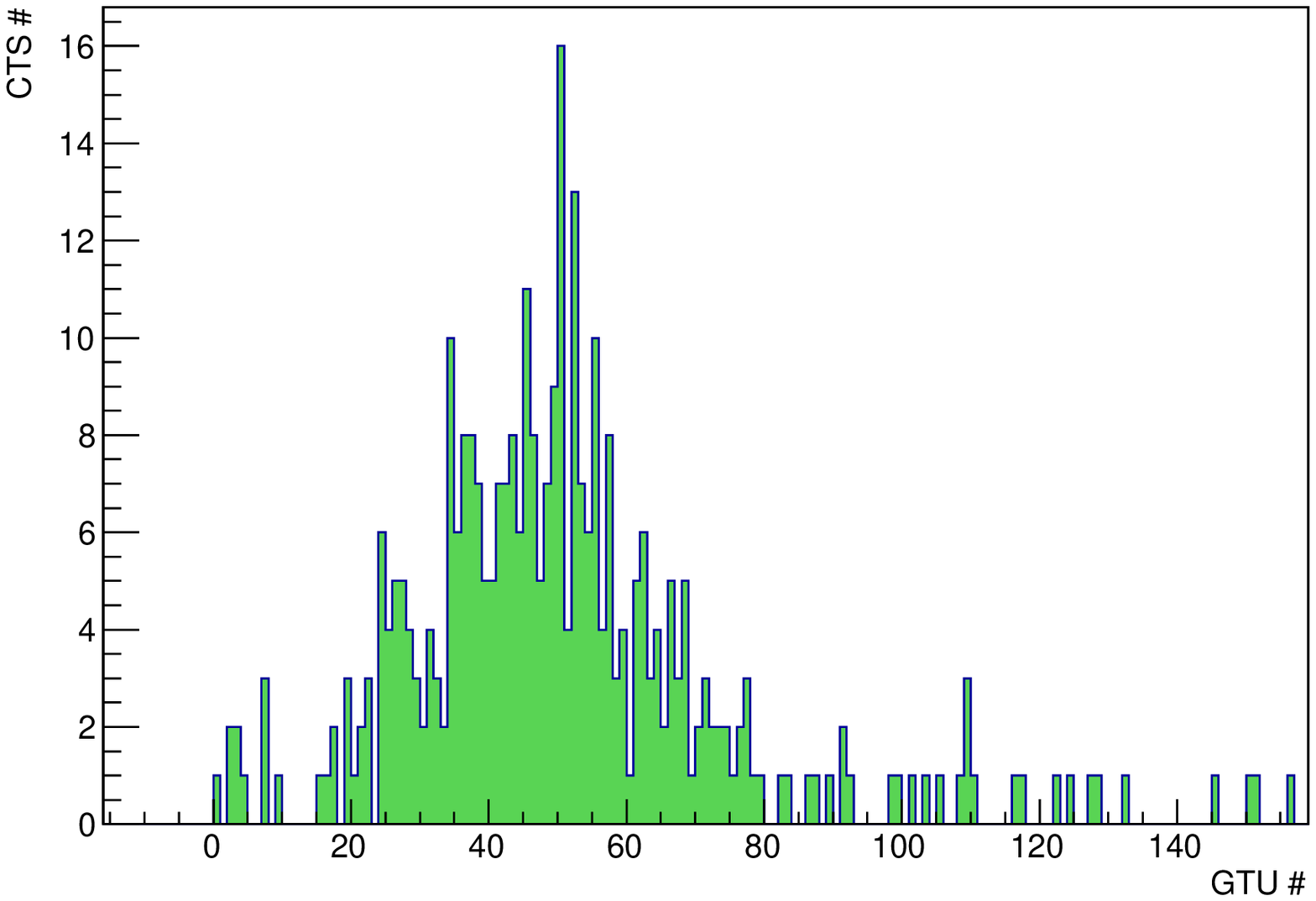}
  \caption{ 
Top: Photon counts observed in the {\me} focal surface for a simulation of an $E$ = \SI{1e21}{\electronvolt} event with an inclination of \SI{80}{\degree} to the nadir. Bottom: Light curve for the same event. The x-axis shows time in units of GTU (1 GTU = \SI{2.5}{\micro\second}). Background of 1~photon~count/pixel/GTU is not included in the simulation, in order to clearly show the track structure. Whilst it is not expected that Mini-EUSO will detect an event of such high energy, this event gives a fair representation of the signal that is expected to be seen by {\me} when the ground-based laser system is used to test the trigger logic, thus verifying the EUSO detection principle.}
  \label{fig:eecr}
  \end{figure}

Although {\me} is not designed to detect EECRs due to the small size of the optical system, it is still possible to detect cosmic rays above the energy threshold of $E_{thr}\sim$ \SI{1e21}{\electronvolt} (see Figure~\ref{fig:trigeff}). Existing results from both ground-based facilities Telescope Array and the Pierre Auger Obsevratory show that we should not expect to detect EECRs at such high energies due to the observation of the GZK suppression at $(5.4~\pm~0.6)~\times~10^{19}$~\si{\electronvolt} and $(2.9~\pm~0.2)~\times~10^{19}$~\si{\electronvolt} respectively \citep{AbuZayyad:2013hg, Abraham:2010iw}. In this way, {\me} will likely provide an upper limit for a null detection with its large annual exposure of $\sim$~\SI{15000}{\kilo\metre\squared\steradian}. This value of the exposure assumes an observational duty cycle of 20\%, as well as effects due to clouds and localised light sources in the FoV. The calculation follows the method presented for JEM-EUSO in \citet{AdamsJr:2013bd}. Figure~\ref{fig:eecr} shows the expected track (top) and light curve (bottom) of a EECR with energy $E$~=~\SI{1e21}{\electronvolt} and inclination of \SI{80}{\degree} to the nadir that would be triggered by the first level trigger of {\me} in standard UV nightglow illumination of $\sim$~1~photon~count/pixel/GTU \citep{AdamsJr:2013bd}. As shown in Figure \ref{fig:trigeff}, such an event is triggered with an efficiency of $\sim$~40\%, and thus is on the threshold of detection. Background is not included in the figure in order to show the structure of the EECR signal, but is implemented in the simulation chain to properly test the trigger logic.

Additionally, during flight it will be possible to simulate EECR-like signals using ground-based laser facilities in order to verify the capability of {\me} to detect cosmic rays and to allow the testing and optimisation of the trigger system. The GLS (Global Light System) has been developed for the JEM-EUSO project in order to provide a means of in-flight calibration via benchmark optical signatures with known rate, intrinsic luminosity, time and direction \citep{AdamsJr:2014fv}. The GLS laser prototype is currently in operation: a portable, steerable, \SI{90}{\milli\joule} and \SI{355}{\nano\metre} laser system based at the Colorado School of Mines. It has been successfully tested both during the flight of EUSO-Balloon \citep{Eser:2015tum} and more recently in the calibration of the EUSO-TA and EUSO-SPB instruments. Even with one fixed, ground-based station, a laser campaign during the ISS overpass is feasible, with an opportunity around once every 2 months during the flight of Mini-EUSO. Simulations show that a \SI{80}{\milli\joule}, unpolarised laser, fired perpendicular to the ISS motion with an elevation angle of \SI{25}{\degree} would produce a signal comparable in both intensity and duration to an EAS produced by an EECR of $\sim10^{21}$~\si{\electronvolt} (see Figure \ref{fig:eecr}).

Other scientific objectives of {\me} include observation of the bioluminescence of the sea from space. Since 1915, there have been 255 documented reports of \emph{milky sea} \citep{great1993marine} and even more events have been reported historically. The \emph{milky sea} or \emph {mareel} is a term used to describe conditions where large areas of the ocean surface (up to \SI{16,000}{\kilo\metre\squared}) appear to glow during the night for periods of up to several days. The condition is poorly understood, but typically attributed to the bioluminescence of the luminous bacteria \emph{Vibrio harveyi} in connection with the presence of colonies of the phytoplankton \emph{Phaeocystis}. The bioluminescent bacteria have been shown in the laboratory to have an emission spectra which peaks at \SI{490}{\nano\metre} with a bandwidth of \SI{140}{\nano\metre} \citep{Hastings:1991wu, Seliger:Rtjyriki}.  There has been a single report of satellite observations of this phenomenon, confirmed by a ship-based account \citep{Miller:2005di}. Whilst the BG3 filter on the {\me} MAPMTs is optimised for the 300~-~400~\si{\nano\metre} band, it extends up to \SI{500}{\nano\metre} and Mini-EUSO is able to detect $\sim$~20\% of the bioluminescence spectrum. Taking this into account, for a signal of 5$\sigma$ above the background level of 1~photon~count/pixel/GTU \citep{AdamsJr:2013bd} in a single pixel of {\me}, the limiting source radiance of the bacteria is $\sim10^{10}$~photons/\si{\centi\metre\squared}/\si{\second}. This number should be regarded as approximate as the true sensitivity also depends on the spatial extent of the signal on the focal plane and the background level, which is dependent of the atmospheric conditions at the time of observation. 5$\sigma$ in a single pixel is stringent requirement for a signal that is expected to cover a significant portion of the focal surface for a duration of around \SI{20}{\second}. The response of {\me} is included, and atmospheric attenuation has been neglected. This estimate gives an order of magnitude higher sensitivity than the value of $1.4\times10^{11}$~photons/\si{\centi\metre\squared}/\si{\second} reported in \citet{Miller:2005di}, following a successful detection. Further detections of the milky sea events from space could deeply enhance the understanding of this elusive phenomena, as well as the distribution and transport of phytoplankton on a global scale.  

\subsection{Technological objectives}
In addition to the scientific objectives, {\me} will address important technology issues regarding the future of EECR detection from space. {\me} will be the first use of a Fresnel-based optics system in space and will provide the opportunity to validate the JEM-EUSO observation scheme on the scale of one module. The technology readiness level of the JEM-EUSO instrumentation will also be raised by this mission providing important spaceflight qualification and heritage of the hardware.

\section{Simulations of typical observations}
The {\me} configuration has been included in the ESAF (EUSO Simulation and
Analysis Software) package. ESAF is the official software tool to perform simulations of EAS development, photon production and transport through the atmosphere and detector response for optics and electronics. Moreover, ESAF includes algorithms for the reconstruction of the properties of air showers produced by EECRs. Originally developed for the ESA-EUSO mission, all the planned missions of the JEM-EUSO program have been implemented in ESAF in order to assess the full range of expected performances for cosmic ray observation \citep{asr}. The simulation and modeling of EECR events in ESAF is detailed in \citet{esaf}. All simulated data shown in Section \ref{sec:science} has been generated using ESAF. 

\begin{table}[h]
\centering
\caption{Typical key parameters used in the modelling of TLEs implemented in ESAF. The altitude, radius and extension all refer to the maximum values reached in the development of the simulation. For blue jets, the angle of the jet with respect to the vertical can also be defined and a typical value is \SI{15}{\degree}. Faint TLEs are simulated here in order to test the sensitivity of {\me} to such events.}
\vspace{3mm}
   \begin{tabular}{ cccccc }
    \toprule
 & \textbf{Altitude} & \textbf{Abs} & \textbf{Radius}  & \textbf{Extension} & \textbf{Duration} \\ 
 & \textbf{[km]} & \textbf{mag.} & \textbf{[km]} & \textbf{[km]} & \textbf{[ms]} \\ \midrule
     \textbf{Blue Jets} & 60 & 2 &  0.3 & 4 & 15 \\
     \textbf{Sprites} & 80 & 1 &  1 & 2 & 10 \\
     \textbf{Elves} & 80 & 1 & 50 & 1 & 20 \\
        \bottomrule
  \end{tabular}
  \label{table:tle_param}
\end{table}

Recently the simulation of slower events, such as TLEs, meteors and space debris has also been implemented in the ESAF framework. It is possible to simulate 3 different types of TLE: blue jets, sprites and elves. Blue jets are modelled as an expanding cone of light in the atmosphere with the emission spectra dominated by the second positive $N_2$ and the first negative $N_2^{+}$ bands as detailed in \citet{Pasko:2002by}. Sprites are modelled as an expanding light cone with a hemispherical top, and are defined similarly to blue jets. Elves are modelled as larger expanding disks of light with a central hole and a spectral profile as described in \citet{Chang:2010ef}. In each case, key parameters define the size and shape, duration, development and brightness of the TLEs. Typical values for these parameters are shown in Table \ref{table:tle_param}. Meteors are modelled using a simple simulator which allows the specification of the initial meteor altitude, the velocity vector, the event duration and the morphology of the light curve due to meteoroid ablation. The possibility of simulating flares in the emission of the meteor during its passage through the atmosphere is also implemented, with flare start time, duration and light curve morphology being further input parameters. More details of the approach to meteor simulation can be found in \citet{PSS}. The modelling of space debris is currently under development. Debris are modelled as spherical and implemented in a similar manner to that of meteors, but taking into account the geometry of the illumination and the kinematics in the field of view.

\section{Conclusion}
In summary, {\me} is a compact UV telescope that will be placed at a nadir-facing window inside the Zvezda module of the ISS. The instrument employs a multi-level trigger system. This allows it to capture interesting events on the time scales of EECR-induced atmospheric showers and TLEs with a high temporal resolution, whilst also providing a continuous readout with a resolution of \SI{40.96}{\milli\second}. {\me} will provide insight into a variety of atmospheric and terrestrial UV phenomena (with complementary information from the NIR and visible light cameras) as well as raising the technology readiness level of future EUSO missions. Typical observations have been simulated in order to verify the mission goals and to test the trigger algorithm. {\me} is approved as a joint project by the Italian (ASI) and Russian (Roscosmos) space agencies and the instrument integration is currently at an advanced stage to be on schedule with a possible launch in late 2017 to early 2018.

\FloatBarrier
\section*{Acknowledgements}
This work was partially supported by the Italian Ministry of Foreign Affairs and International Cooperation,  Italian Space Agency (ASI) contract 2016-1-U.0, the Russian Foundation for Basic Research, grants \#15-35-21038 and \#16-29-13065, and the Olle Engkvist Byggm{\"a}stare Foundation. We acknowledge useful discussions with M.~Bertaina and F.~Fenu regarding the ESAF simulations and also the contribution of Y.~Takizawa in the simulation of the Mini-EUSO optical system. The anonymous referees are also thanked for their detailed and constructive input. The authors would like to dedicate this paper to the memory of Dr. Jacek Karczmarczyk and Dr. Yoshiya Kawasaki, who have contributed greatly to the collaboration and will be deeply missed. 

\section*{References}

\bibliography{mybibfile}

\end{document}